\documentclass[11pt]{article}
\usepackage{amsmath,amsfonts}
\usepackage{amssymb}
\usepackage{latexsym}
\def \beq  {\begin{equation}}
\def \eeq  {\end{equation}}
\def \beqar {\begin{eqnarray}}
\def \eeqar {\end{eqnarray}}
\def\sqr#1#2{{\vcenter{\vbox{\hrule height.#2pt
\hbox{\vrule width.#2pt height#1pt \kern#1pt
\vrule width.#2pt}\hrule height.#2pt}}}}

\def\t0{\tilde{0}}

\begin{document}
\def \CMP {{Commun. Math. Phys.}}
\def \PRL {{Phys. Rev. Lett.}}
\def \PL {{Phys. Lett.}}
\def \NPBProc {{Nucl. Phys. B (Proc. Suppl.)}}
\def \NP {{Nucl. Phys.}}
\def \RMP {{Rev. Mod. Phys.}}
\def \JGP {{J. Geom. Phys.}}
\def \CQG {{Class. Quant. Grav.}}
\def \MPL {{Mod. Phys. Lett.}}
\def \IJMP {{ Int. J. Mod. Phys.}}
\def \JHEP {{JHEP}}
\def \PR {{Phys. Rev.}}
\def \JMP {{J. Math. Phys.}}
\def \GRG{{Gen. Rel. Grav.}}
\fontfamily{cmr}\fontsize{11pt}{17.2pt}\selectfont
\begin{titlepage}
\null\vspace{-62pt} \pagestyle{empty}
\begin{center}
\rightline{CCNY-HEP-08/5}
\rightline{November 2008}
\vspace{1truein} {\Large\bfseries
Massive Spin-2 fields of Geometric Origin in Curved}\\
\vskip .1in
{\Large\bfseries Spacetimes}\\
\vspace{.5in} {\large V. P. NAIR$^a$, S. RANDJBAR-DAEMI$^b$,
V. RUBAKOV$^c$}~\footnote{E-mail:
{\fontfamily{cmtt}\fontsize{11pt}{15pt}\selectfont
vpn@sci.ccny.cuny.edu, seif@ictp.trieste.it, rubakov@ms2.inr.ac.ru}}\\
\vspace{.1in}{\itshape $^a$Physics Department, City College of the
CUNY, New York, NY 10031}\\
\vspace {.1in}{\itshape $^b$The Abdus Salam International Centre
for Theoretical Physics, Trieste, Italy}\\
\vspace{.1in}{\itshape $^c$Institute for Nuclear Research of the Russian Academy of
Sciences,
Moscow, Russia}\\

\vspace{.4in}
\centerline{\large\bf Abstract}
\end{center}
We study the consistency of a model which includes torsion as well as
the metric as dynamical fields and has massive spin-2 particle in
its spectrum. 
It is known that this model
is tachyon- and ghost-free in Minkowski background. We show that
this property remains valid and no other pathologies emerge
in de~Sitter and anti-de~Sitter
backgrounds, with some of our results extending to arbirary
Einstein space backgrounds. This suggests that the model is
consistent, at least at the classical level, unlike,
e.g., the Fierz--Pauli theory.

\end{titlepage}
\pagestyle{plain} \setcounter{page}{2}

\section{Introduction and summary}
\label{sec:intro}
The aim of this note is to examine the possibility of obtaining a
geometrical field theory, consistent
at least at the classical level, for an
interacting massive spin-2 particle.  The problems with massive spin-2
particles are well illustrated by the Fierz--Pauli
theory~\cite{Fierz:1939ix} of massive graviton.  In Minkowski
background, the latter theory is ghost-free and describes the
propagation of 5 degrees of freedom, the right number for massive
spin-2 particle. One of the problems arising already in Minkowski
background is the van~Dam--Veltman--Zakharov
discontinuity~\cite{vanDam:1970vg,Zakharov:1970cc} and strong coupling
related to it~\cite{Arkani-Hamed:2002sp}. Even worse, once the
background is not Minkowski, a new, Boulware--Deser mode starts to
propagate~\cite{Boulware:1973my,Creminelli:2005qk,Deffayet:2005ys},
and this mode is necessarily a ghost. Yet another problem with
the  Fierz--Pauli
theory is that even otherwise healthy modes exhibit superluminal
propagation~\cite{Osipov:2008dd}.

Some of these problems may be cured by considering theories whose
``natural'' backgrounds are not
Minkowski~\cite{Deser:2001pe,Deser:2001wx,Porrati:2001db} or
theories with broken
Lorentz-invariance~\cite{Rubakov:2004eb,Dubovsky:2004sg,
Dubovsky:2004ud,Berezhiani:2007zf} (for a review see, e.g.,
Ref.~\cite{Rubakov:2008nh}). On the other hand,
there is another class of classical theories which are known to admit
 consistent propagation of both massless
gravitons
and massive spin-2 modes in Minkowski
 background~\cite{Hayashi:1979wj,Hayashi:1980ir,Hayashi:1980qp,Sezgin:1979zf}.
These theories treat the vierbein and the
 connection as independent variables and include, in addition to the
 scalar curvature of the connection, quadratic terms in the full
 curvature tensor and its contractions. They also include bilinear
 terms in the torsion which eventually act like mass terms for
 propagating modes emerging from the torsion. It is a theory from this
class that we focus on in this note; we will specify the Lagrangian of the
theory below in this Section. Our purpose is to study whether this theory
exhibits the Boulware--Deser phenomenon and/or superluminal
propagation in curved backgrounds.

We note that there is a subclass of theories studied in 
Refs.~\cite{Hayashi:1979wj,Hayashi:1980ir,Hayashi:1980qp,Sezgin:1979zf}
in which, besides massless graviton, only spin-0 modes propagate
in Minkowski background. It has been 
argued~\cite{Hecht:1996ay,Yo:1999ex} that the models from this subclass
are fully consistent. This is in agreement with the expectation that
a theory whose spectrum about Minkowski background
does not include  massive spin-2 states may not be problematic
in curved backgrounds. Unlike in Refs.~\cite{Hecht:1996ay,Yo:1999ex},
we study here the theory that does contain propagating massive spin-2
modes in Minkowski background, and thus, off hand, has less chance to be
fully consistent.

In the Fierz--Pauli theory, the Boulware--Deser mode and superluminal propagation
show up already in backgrounds of the highest possible symmetry admitted in
that theory~\cite{Rubakov:2008nh,Osipov:2008dd}. Thus, as the first step,
it is natural to study what are the most symmetric backgrounds admitted by the
theory we are interested in, and then analyze the propagation of perturbations
about these backgrounds.
We find that the theory admits Einstein spaces
with zero torsion as solutions to the field equations. These include the
maximally symmetric, de~Sitter and anti-de~Sitter spaces. Our main result is that
the number of propagating modes in these maximally symmetric
backgrounds is exactly the same as in Minkowski space, and that
these modes obey the standard massive equations of the Klein--Gordon type.
Since all propagating modes are not ghosts in Minkowski background,
they are not ghosts in the curved backgrounds we study. The latter
property follows from continuity argument: as the
parameters of our model change continuously, the background
smoothly interpolates between Minkowski and (anti-)de~Sitter spaces
of different curvature, while the kinetic terms in the field equations
do not become singular or vanish; hence the kinetic part of the action
for propagating modes does not change sign.
Thus, there is neither Boulware--Deser phenomenon nor superluminal propagation,
at least in maximally symmetric backgrounds. This suggests that the theory
we study in this note, and possibly other theories from the class described in
Refs.~\cite{Hayashi:1979wj,Hayashi:1980ir,Hayashi:1980qp,Sezgin:1979zf},
may be consistent theories of massive spin-2 particles.
It is worth pointing  out that these theories have been
considered in an astrophysical context with constraints on the
parameters of the Lagrangian from observational data~\cite{Mao:2008gv}.

To describe the class of  models of
Refs.~\cite{Hayashi:1979wj,Hayashi:1980ir,Hayashi:1980qp,Sezgin:1979zf},
 we  denote the vierbein as usual by $e_\mu^i$ and the
 connection, which can be regarded as an $O(1,3)$ gauge field, by
 $A_{ij\mu}=-A_{ji\mu}$, where $\mu=0,1,2,3$ is the space-time index and
 $i,j=0,1,2,3$ are the tangent space indices. The curvature is then
 defined as in any Yang-Mills theory, namely\footnote{In this paper we
 closely follow the notations of Refs.~\cite{Hayashi:1979wj,
 Hayashi:1980ir, Hayashi:1980qp}.}
\beq
 F_{ijmn}=e_{m}^{\mu}e_{n}^{\nu}(\partial_\mu A_{ij\nu}-\partial_\nu
 A_{ij\mu} +A_{ik\mu}{A^{k}}_{j\nu}-A_{ik\nu}{A^{k}}_{j\mu})
\nonumber
\eeq
Throughout this paper we  mostly use the tangent space basis.
In this basis the indices are
raised and lowered by the Minkowski metric $\eta_{ij}$
(signature is mostly positive).

From the vierbein $e_\mu^i$ and its inverse $e_i^\mu$ one
constructs an object denoted by $C_{ijk}$ and defined
by
\beq C_{ijk}=e_j^\mu e_k^\nu(\partial_\mu
e_{i\nu}-\partial_\nu e_{i\mu})
\label{C}
\eeq
This object together with the connection $A_{ijk}=e_{k}^\mu
A_{ij\mu}$ enables one to introduce  the torsion tensor
$T_{ijk}$,
\beq T_{ijk}=
A_{ijk}-A_{ikj}-C_{ijk}
\nonumber
\eeq
The actions studied in this paper will be a subclass of the most
general actions which are quadratic in $T_{ijk}$ as well as in
$F_{ijkl}$ and its all possible contractions with $\eta_{ij}$ and
$\varepsilon_{ijkl}$. These contractions are
\beq
F_{jl}=
\eta^{ik}F_{ijkl},\quad\quad\quad F= \eta^{jk}F_{jk},\quad\quad\quad
\varepsilon \cdot F= \varepsilon_{ijkl}F^{ijkl}
\label{contract}
\eeq
If $T_{ijk}$ vanishes, then $F_{ijkl}, F_{ij}$ and $F$ reduce
respectively to the Riemann curvature tensor, Ricci tensor and Ricci
scalar. In that case one has $F_{ij}=F_{ji}$, otherwise this tensor
generically
has both symmetric and antisymmetric parts. Also, $F_{ijkl}$ is
antisymmetric with respect to the interchange $i\leftrightarrow j$ and
$k\leftrightarrow l$. Unlike the Riemann tensor of the standard
torsion free connection, $F_{ijkl}$ is not symmetric with respect to
the interchange of the first pair of its indices with the last
pair.

The tensor $T_{ijk}= -T_{ikj}$ can be decomposed into its irreducible
 components under the action of the local $O(1,3)$
 transformations. They are given by
\beq
 T_{ijk}=\frac{2}{3}(t_{ijk}-t_{ikj})+\frac{1}{3}(\eta_{ij}v_k
 -\eta_{ik}v_j)+\varepsilon_{ijkl}a^l
\label{T'}
\eeq
where the field $t_{ijk}$ is symmetric with respect to the
interchange of $i$ and $j$ and  satisfies the cyclic and trace
identities,
\beq
t_{ijk}+t_{jki}+t_{kij}=0, \quad\quad\quad
\eta^{ij}t_{ijk}=0,\quad\quad\quad \eta^{ik}t_{ijk}=0
\nonumber
\eeq
Due to the cyclic identity, all components of $t_{ijk}$ can
be expressed in terms of the antisymmetric part $t_{i[jk]}$,
\beq
t_{ijk}= \frac{2}{3}\left(t_{i[jk]}+t_{j[ik]}\right)
\label{dec19-1}
\eeq
The antisymmetric tensor  $t_{i[jk]}$ also obeys the
cyclic and trace identities,
\beq
t_{i[jk]}+t_{j[ki]}+t_{k[ij]}=0, \quad\quad\quad
\eta^{ij}t_{i[jk]}=0 \; .
\nonumber
\eeq
Making use of the latter property, one finds that $t_{i[jk]}$, and hence
$t_{ijk}$, has 16 independent components\footnote{To see this, one notices
that the cyclic identities
for $t_{i[jk]}$ are trivially satisfied if any two of the indices
are equal to each other. Hence, there are $4!/3! =4$ non-trivial
cyclic identities. Trace identities, instead, involve
$t_{i[jk]}$ with two  indices equal to each other, and there are 4 such identities.
Hence, 8 out of 24 components of $t_{i[jk]}$ are not independent,
which leaves 16 independent components.}. Together with 4 components of the
vector $v_i$ and 4 components of pseudo-vector $a_i$ these make 24 independent
components of $T_{ijk}=-T_{ikj}$, as they should.

Now we have all the algebraic preliminaries to define the general
class of the actions of interest  to us.  The action is given by
$S=\int d^4x ~ e ~L$, where $e= \det e_\mu^i$ and $L=L_F+ L_T$, 
with\footnote{The mass terms in $L_T$ can also be motivated
 in terms of spontaneous breaking of the gauge symmetry 
$SO(1,3)$~\cite{Percacci:1990wy}.}
\beq
L_F= c_1F + c_2 +
c_3F_{ij}F^{ij}+c_4F_{ij}F^{ji}+c_5F^2+c_6(\varepsilon_{ijkl}F^{ijkl})^2+
bF_{ijkl}F^{ijkl},
\label{LF0}
\eeq
\beq
L_T= \alpha t_{ijk}t^{ijk} +\beta v_iv^i+ \gamma a_ia^i
\label{LT0}
\eeq
Note that an extra term of the form $F_{ijkl}F^{klij}$ can also be
included in (\ref{LF0}).
However, using the 4-dimensional Gauss-Bonnet
invariant, this term can be expressed through the other quadratic
curvature invariants already included in (\ref{LF0}).  
The defining property of the class of theories (\ref{LF0}), (\ref{LT0})
is that the derivative terms are geometric (they are expressed through
the curvature tensor) while non-derivative terms are quadratic in
torsion; the expression (\ref{LF0}) is the most general quadratic
expression
(assuming parity conservation).

For the special
case of $c_2=0$ and in  Minkowski background,
the spectrum of propagating modes has been studied in detail in
Refs.~\cite{Hayashi:1979wj,Hayashi:1980ir,Hayashi:1980qp,Sezgin:1979zf}.
These studies have shown that in order to obtain a tachyon- and
ghost-free perturbation spectrum about Minkowski space, the parameters
of this action should be restricted in some specific manner. In this
way all possible tachyon- and ghost-free theories have been
tabulated. Among various possibilities there is a class of models
which have, in addition to the massless graviton, a massive spin-2
propagating degree of freedom. This massive spin-2 mode originates from
the $t_{ijk}$-component of the torsion tensor.  In this note we
consider a model belonging to the latter class.
In
Minkowski
background this model
has  massless graviton,  massive spin-2 mode as well as
massive spin-0 propagating degree of freedom.  This is achieved by
setting
$b=c_3+c_4+3c_5=0$ and $\alpha=-\beta=\frac{4\gamma}{9}$.
The parameter $c_2$ has been set equal to zero in
Refs.~\cite{Hayashi:1979wj,Hayashi:1980ir,Hayashi:1980qp,Sezgin:1979zf}
but we  keep it different from
zero.
This will enable us to have (anti-)de~Sitter space as a solution to
the field equations.
Our Lagrangian  $L=L_F+ L_T$ is therefore  given by
\beq
L_F= c_1F + c_2 +
c_3F_{ij}F^{ij}+c_4F_{ij}F^{ji}+c_5F^2+c_6(\varepsilon_{ijkl}F^{ijkl})^2,
\nonumber
\eeq
\beq
L_T= \alpha \left(t_{ijk}t^{ijk} -v_iv^i+\frac{9}{4} a_ia^i \right) \; ,
\nonumber
\eeq
 where the only restriction is
\beq
c_3+c_4+3c_5=0 \; .
\nonumber
\eeq
We will show that the properties described above --- the
Einstein spaces as solutions to the field equations and the
absence of both the Boulware--Deser phenomenon and superluminal propagation
in maximally symmetric spaces
--- hold without any further restrictions on the parameters
$c_1, \dots, c_6$ and $\alpha$ entering the Lagrangian beyond what is required in flat background
\footnote {For the class of models we consider in this paper  the absence of ghosts and tachyons in the flat background  require that $c_5<0, c_6>  0, \alpha<0 $, and $\tilde\alpha=\alpha +\frac{2}{3}c_1 >0$. Note that $c_1=M_p^2$.}.
The masses of the spin-2 and spin-0 modes will pick
up contributions due to the cosmological constant $c_2$
and will remain non-tachyonic if they are non-tachyonic in Minkowski
space.
 except for a stronger bound on $c_5$ in the de Sitter background.  
In order for the spin 
zero particle to be non tachyonic in  
de Sitter space $c_5$ will have to be 
bounded from above. This bound reduces to that of the flat space 
bound when the curvature vanishes.  
This is discussed in section 5.

The rest of this paper is organized as follows.  In section 2 we present
the full non-linear field equations in the model we study. In section 3
we show that torsion-free Einstein spaces are solutions to these
equations.
In section 4 we analyze the perturbations associated with fields $v^i$
and $a^i$, about arbitrary Einstein backgrounds. In section 5 we
study the perturbations originating from the field $t_{ijk}$, about
maximally symmetric spaces. We conclude in section 6.

\section{Field equations}

In the first place, let us present the field equations in our model.
We begin with the gravitational field equations.

\subsection{Gravitational field equations}

The gravitational field equations are found, as usual, from $\frac{\delta
S}{\delta e_m^i}=0$. This leads to
\beq c_1F_{ji}+ c_3(F^m{_i}F_{mj} -
{{F_{j}}^{mn}} _{i} F_{mn})+ c_4(F^m{_i}F_{jm} - {{F_{j}}^{mn}} _{i}
F_{nm})+ 2c_5F_{ji}F\nonumber \eeq
\beq
+2c_6\varepsilon_{kmnj}{F^{kmn}}_i(\varepsilon_{rpqs}F^{rpqs})
+(D^k+v^k)F_{ijk}+ H_{ij}-\frac{1}{2}\eta_{ij}L=0
\label{Ein}
\eeq
where \beqar F_{ijk}&=&
\alpha\left[(t_{ijk}-t_{ikj})-(\eta_{ij}v_k
-\eta_{ik}v_j)-\frac{3}{4}\varepsilon_{ijkl}a^l\right]
\nonumber\\
H_{ij}&=&T_{mni}{F^{mn}}_j-\frac{1}{2}T_{jmn}{F_i}^{mn}
\nonumber
\eeqar
and $D_i$ is the covariant derivative with respect to the
connection $A_{ijk}$.  Note that $H_{ij}$ is second order in the
torsion, therefore, in torsion-free backgrounds it will not
contribute to the linearized equations for perturbations.
Since we are going to consider only such backgrounds,
this tensor will not play any role in our calculations.  Nevertheless
we  keep it for the time being. We also note that the above
equations have both symmetric and antisymmetric parts.

The antisymmetric part of the gravitational field equations
(\ref{Ein}) is straightforward to obtain. It reads
\beqar &&c_1 F_{[ji]}
+\frac{c_4}{2}(F^m{_i}F_{jm} - {{F^{m}}_{j}} F_{im})
-\frac{1}{2}({{F_j}^{mn}}_i-{{F_i}^{mn}}_j)(c_3F_{mn}+c_4F_{nm})\nonumber\\
&&\hskip .2in + 2c_5F_{[ji]}F+c_6(\varepsilon_{kmnj}{F^{kmn}}_i-
\varepsilon_{kmni}{F^{kmn}}_j)(\varepsilon_{rpqs}F^{rpqs}) \nonumber\\
&&\hskip .4in+(D^k+v^k)F_{[ij]k}+ H_{[ij]}=0 \; .
\label{AntEin}
\eeqar
Taking the trace of eq.~(\ref{Ein})
gives  a
useful constraint on $F$, namely,
\beq c_1F= -3\alpha \nabla_iv^i
-L_T-2c_2
\nonumber
\eeq
where $\nabla_i$ is the covariant derivative with respect to the
spin connection $\omega^i_{j\mu}$ derived from the vierbein
$e_\mu^i$.
Let us point out that for
a torsion-free
solution to the field equations, $F$ is a constant given by
\beq
F=-\frac{2c_2}{c_1} \; .
\label{6*-jul}
\eeq
For
later use we also record the linear part of $F$ in a torsion-free
background,
\beq c_1F_{(1)}= -3\alpha \nabla_iv^i
\nonumber
\eeq
We do not use special notation for the background
objects, unless there is a risk of an ambiguity;
the subscript $(1)$ refers to linearized perturbations.
For torsion-free backgrounds discussed in this paper,
the torsion components $v^i$, $a^i$ and $t_{ijk}$ are
perturbations by themselves; we do not label them by
the subscript $(1)$.
 Hereafter
the
operator $\nabla_i$
always denotes the covariant derivative with respect to the
background metric.

Another useful formula
is the
first order relation between $F_{(1)}$ and the scalar curvature
$R_{(1)}$ in a torsion-free background, where $R_{(1)}$ is
derived from the spin connection $\omega^i_{j\mu}$.
By making use of the definition of $F$ and expanding in veirbein
perturbations and
torsion, one finds
\beq F_{(1)}= R_{(1)}+ 2\nabla_iv^i
\nonumber
\eeq
Combining the last two equations we obtain the first order
relation between $R_{(1)}$ and $v$ in a torsion-free background,
namely,
\beq
c_1R_{(1)}=-3\tilde\alpha \nabla_iv^i
\nonumber
\eeq
where $\tilde\alpha$ is a useful parameter which will be
encountered frequently in the sequel,
\beq
3\tilde\alpha= 2c_1+ 3\alpha
\nonumber
\eeq

\subsection{Torsion field equations}

We now move on to the torsion field equations.
These
are obtained from $\frac{\delta S}{\delta
A^i_{j\mu}}=0$. Written in the orthonormal basis they become
\beqar
&&c_3\left\{\eta^{ik}(D_m+\frac{2}{3}v_m)F^{jm}-\eta^{jk}(D_m+\frac{2}{3}v_m)F^{im}-(D^i+\frac{2}{3}v^i)F^{jk}+(D^j+\frac{2}{3}v^j)F^{ik}\right\}\nonumber\\
&&+c_4\left\{\eta^{ik}(D_m+\frac{2}{3}v_m)F^{mj}-\eta^{jk}(D_m+\frac{2}{3}v_m)F^{mi}-(D^i+\frac{2}{3}v^i)F^{kj}+(D^j+\frac{2}{3}v^j)F^{ki}\right\}\nonumber\\
&&+c_5\left\{\eta^{ik}(D^j+\frac{2}{3}v^j)F-\eta^{jk}(D^i+\frac{2}{3}v^i)F\right\}+4c_6\left\{\varepsilon^{ijkm}(D_m+\frac{2}{3}v_m)(\varepsilon \cdot F)\right\}\nonumber\\
&&-\left(\frac{4}{3}{t^k}_{[mn]}
+{\varepsilon^k}_{mnp}a^p\right)\left\{\phantom{\frac{}{}}  c_3(\eta^{im}F^{jn}-\eta^{jm}F^{in})+
c_4(\eta^{im}F^{nj}-\eta^{mj}F^{ni})\right. \nonumber\\
&&\left.+2c_5\eta^{im}\eta^{jn}F+2c_6\varepsilon^{ijkm}(\varepsilon \cdot F)\right\}+H^{ijk}=0
\label{torsion}
\eeqar
where
\beq
H_{ijk}= -\tilde\alpha(t_{kij}-t_{kji})+\tilde\alpha(\eta_{ki}v_j -\eta_{kj}v_i)-\frac{3\tilde\alpha}{2}\varepsilon_{ijkl}a^l
\nonumber
\eeq
Equations (\ref{torsion}) can be decomposed exactly in the same
manner as the torsion itself has been decomposed into $t_{ijk}, v_i$
and $a_i$. 
This is done by taking the trace of eq.~(\ref{torsion}) over $j$
and $k$ and by contracting eq.~(\ref{torsion}) with
$\varepsilon_{ijkl}$. In the course of these manipulations we
make use of the constraint $c_3+c_4=-3c_5$.

The trace of the torsion equation is obtained by
contracting eq.~(\ref{torsion})
with $\eta^{jk}$, 
\beqar
&&-3c_5\left( D_jF^{(ij)}-\frac{1}{2}D^iF\right)-2c_5\left(V_jF^{(ij)}-\frac{1}{2}V^iF\right)+(c_3-c_4)D_jF^{[ij]}
\nonumber\\
&&+\frac{2}{3}(c_3-c_4)V_iF^{[ij]}-3c_5t^{i(jn)}F_{(jn)}+\frac{1}{3}(c_3-c_4)t^{i[jn]}F_{[jn]}
\nonumber\\
&&-\frac{1}{2}(c_3-c_4)\varepsilon^{ijnl}a_lF_{[jn]}+6c_6a^i(\varepsilon \cdot F) + \frac{3}{2}\tilde\alpha v^i=0
\label{trtorsion}
\eeqar
The curl of the torsion equation
is found
by contracting eq.~(\ref{torsion}) with $\varepsilon_{ijkl}$,
\beqar
&&(c_3-c_4)\varepsilon_{lijk}D^iF^{jk} -12c_6 D_l(\varepsilon \cdot F)-\frac{2}{3}\varepsilon_{ijkl}{t_n}^{ik}(c_3F^{jn}+c_4F^{nj})-8c_6v_l (\varepsilon \cdot F)
\nonumber\\
&&-\frac{2}{3}(c_3-c_4)\varepsilon_{ijkl}v^iF^{jk}-2(c_3F_{jl}+c_4F_{lj})a^j +\frac{2}{9}\tilde\alpha a_l=0
\label{curltorsion}
\eeqar

\section{Backgrounds}

As mentioned above, we  consider torsion-free backgrounds only.
In this case $F_{ijkl}=R_{ijkl}, F_{ij}=R_{ij}$ and $F=R$, where
$R_{ijkl},R_{ij}$ and $R$ denote, respectively, the Riemann tensor,
the Ricci tensor and the Ricci scalar of the background metric.

The
gravitational equations (\ref{Ein}) reduce to
\beq
c_1R_{ij}-3c_5(R{^m}_iR_{mj} -{{R_j}^{mn}}_i R_{mn})- 4\frac{\lambda
c_5}{c_1}R_{ij} -\frac{1}{2}\eta_{ij}L=0
\label{grav}
\eeq
where
\beq L= -3c_5 R_{ij}R^{ij} +
4c_5\left(\frac{c_2}{c_1}\right)^2 -c_2
\nonumber
\eeq
These equations should be
solved alongside with the torsion equation (\ref{torsion}) which
yields
\beq \nabla_iR_{jk}-\nabla_jR_{ik}=0
\label{co}
\eeq
In writing these equations we made use of the fact that 
scalar curvature of the
background manifold is  constant, see (\ref{6*-jul}). 
Combining eqs.~(\ref{grav}) and
(\ref{co})
with the
Bianchi identity we obtain a condition on the Riemann tensor or
equivalently on the Weyl tensor of the background manifold, namely,
\beq
\nabla^i R_{ijkl}=0=\nabla^i W_{ijkl}
\nonumber
\eeq
The Weyl
tensor is defined by
\beq R_{ijkl}= W_{ijkl}
+\frac{1}{2}(\eta_{ik}R_{jl}-\eta_{il}R_{jk}
-\eta_{jk}R_{il}+\eta_{jl}R_{ik})-\frac{1}{6}(\eta_{ik}\eta_{jl}
-\eta_{il}\eta_{jk})R
\nonumber
\eeq
The tensor
$W_{ijkl}$ has all the symmetries of the Riemann tensor plus the
additional property that it is trace-free in all pairs of
indices.

By inspecting eqs.~(\ref{grav}) and (\ref{co}) we find that they
 are identically satisfied for  Einstein manifolds (whose definition is
$R_{ij} = \mbox{const} \cdot \eta_{ij}$).
For such manifolds we obtain
\beq
R_{ijkl}=\Lambda(\eta_{ik}\eta_{jl}-\eta_{il}\eta_{jk}) +
W_{ijkl}, \quad \quad R_{ij}=3\Lambda\eta_{ij}, \quad\quad R=
12\Lambda,
\nonumber
\eeq
and the only constraint we arrive at is
\beq
\Lambda = -\frac{c_2}{6c_1} \; .
\label{cc}
\eeq
In this respect our theory, unlike some other theories
whose Lagrangians are of the second order in the Riemann tensor,
is similar to General Relativity.

\section{Linearized theory in Einstein backgrounds}

In this section we study field perturbations about general
torsion-free Einsten backgrounds. We will be able to go pretty
far in our analysis. Namely, we will show that the vector field
$v_i$ does not have its own propagating modes, as it is expressed
through the field $t_{ijk}$. We will also show that the field
$a_i$ is a gradient, and its longitudinal part obeys the massive
Klein--Gordon equation. These properties hold for arbitrary
Einstein space background and are exactly the same as
in the theory about Minkowski background. On the other hand,
to find explicitly the spectrum of perturbations associated with
the field $t_{ijk}$, we will have to resort to maximally symmetric
backgrounds. This is done in section \ref{sec:symmetric}.

\subsection{Consequences of gravitational equations}
\label{sub:cons-grav}

\subsubsection{Antisymmetric gravitational equations}
\label{subsub:anti-grav}

Important constraints are obtained by considering
 antisymmetric part of the gravitational equations.
It is  straightforward  to show that in the backgrounds of
Einstein manifolds eq.~(\ref{AntEin}) reduces to
\beq
(c_1-4\Lambda c_3)F_{(1)[ji]}
-\nabla^kF_{(1)[ji]k}-(c_3-c_4){{W_j}^{[mn]}}_i F_{(1)[mn]}=0 \; .
\label{LinAnt}
\eeq
By linearizing $F_{ij}$ defined in (\ref{contract}) we find
\beqar
 F_{(1)[ji]}&=&-\frac{2}{3\alpha}\nabla^kF_{[ji]k}\nonumber\\
 &=&\frac{2}{3}(\nabla^k t_{k[ji]}-{\nabla_{[j}}v{_i}_]+\frac{3}{4}\varepsilon_{jikl}\nabla^ka^l)
 \label{F1*}
 \eeqar
Hence,
eq.~(\ref{LinAnt}) takes the following form,
 \beq
  ({{I_{i}}^{[mn]}}_j+a{{W_{i}}^{[mn]}}_j)(\nabla^k t_{k[mn]}-{\nabla_{[m}}v{_n}_])=-\frac{3}{4}(\varepsilon_{ijkl} +a {{W_{i}}^{[mn]}}_j\varepsilon_{mnkl})\nabla^ka^l
  \label{25}
\eeq
where
\beqar
a&=&- \frac{\frac{2}{3}(c_3-c_4)}{\alpha+\frac{2}{3}(c_1-4\Lambda c_3)}\label{33a}\\
{{I_{i}}^{[mn]}}_j&=& \frac{1}{2}(\delta^m_i\delta^n_j-\delta^n_i\delta^m_j)
\nonumber
\eeqar
Equation (\ref{25}) is solved by

\beq
\nabla^k t_{k[ij]}-{\nabla_{[i}}v{_j}_]=-\frac{3}{4}\varepsilon_{ijkl}\nabla^ka^l
\label{F1=0}
\eeq
Inserting this in eq.~(\ref{F1*}) we obtain
\beq
\nabla^k F_{[ij]k} = 0
\nonumber
\eeq
and hence
\beq
F_{(1)[ij]}=0
\label{F1=0'}
\eeq
This result plays a crucial role in simplifying all  other equations.

\subsubsection{Symmetric gravitational equations}
 Inserting the results of section \ref{subsub:anti-grav} in the gravitational
 field equations (\ref{Ein}) and linearizing in the Einstein
 background we obtain the equations for the symmetric 
components\footnote{Note that,
 whenever convenient, we drop the symmetrization bracket $(ij)$ from
 $F_{(1)(ij)}$ and $\nabla^kF_{(1)(ij)k}$ as the antisymmetric parts of
 these tensors are zero to the first order.} of
 $F_{(1)(ij)}= F_{(1)ij}$. These equations read
\beq
c_1\left( F_{(1)ij}-\frac{1}{2}\eta_{ij}F_{(1)}
\right) +\nabla^k F_{(1)ijk}+
3c_5W_{jmni}F_{(1)}^{mn}=0
 \label{symmet}
 \eeq
Linearizing  $F_{ij}$ and recalling the definition of $F_{ijk}$ we write
\beqar
F_{(1)ij}&=& R_{(1)ij}-2\nabla^k t_{k(ij)} +\frac{1}{3} (\nabla_{i}v_j+\nabla_{j}v_i)+\frac{1}{3}\eta_{ij}\nabla \cdot v
\label {S}
\\
\nabla^kF_{(1)ijk}&=& -\alpha
\left[3\nabla^k t_{k(ij)} -\frac{1}{2} 
(\nabla_{i}v_j+\nabla_{j}v_i)+\eta_{ij}\nabla \cdot v \right]
\label{S'}
\eeqar
Substitution in (\ref{symmet}) yields
\beq
c_1R_{(1)ij}= -\frac{\tilde\alpha}{2}\eta_{ij}\nabla \cdot v +3\tilde\alpha\nabla^kt_{k(ij)} - \frac{\tilde\alpha}{2}(\nabla{_i}v_j+\nabla{_j}v_i) -3c_5W_{imnj}F_{(1)}^{mn}
\label{R}
\eeq
This equation can be viewed as
the relationship between the gravitational perturbation contained
in $R_{(1)ij}$ and perturbations of the torsion field.
Therefore, eq.~(\ref{R}) is irrelevant for the study of the
spectrum of the torsion perturbations. This spectrum comes out
from the analysis of the torsion field equations.
Of course, the homogeneous part
of eq.~(\ref{R}) describes the propagation of massless gravitons in
the Einstein backgrounds, just like in General Relativity.

\subsection{Linearized torsion field equations}
\subsubsection{The pseudovector $a_l$}
\label{subsub:a}
Substituting (\ref{F1=0'}) in eq.~(\ref{curltorsion}) and using the
fact that $\varepsilon \cdot F= 6\nabla^ia_i$ we obtain the following
linearized equation for $a_l$,
\beq 8c_6\nabla_l(\nabla \cdot a)-
\left( 2\Lambda c_5 +\frac{\tilde\alpha}{2}\right) a_l=0
\label{a}
\eeq
This equation shows that the pseudovector field $a_l$ is a
gradient. As a result the right hand side of (\ref{F1=0}) vanishes
and  eq.~(\ref{F1=0}) becomes the important constraint
\beq \nabla^k
t_{k[mn]}={\nabla_{[m}}v{_n}_]
\label{Const}
\eeq
Furthermore by acting with $\nabla^l$ on (\ref{a}) we obtain the
Klein-Gordon equation for the longitudinal part $\sigma=\nabla^ia_i$
of the pseudovector field,
\beq \left(\nabla^2 -\frac{2\Lambda c_5
+\frac{\tilde\alpha}{2}}{8c_6}\right) \sigma=0
\label{aeq}
\eeq
It can be verified that the mass of the spin zero field $\sigma$
coincides with the flat space result when $\Lambda=0$. Thus, the only
propagating degree of freedom of the field $a_i$ is its longitudinal
part $\sigma$. Its transverse part is simply zero.

\subsubsection{The vector field $v_i$}
\label{subsub:v}
Next we obtain the equation for the vector field $v_i$. To this end we
substitute $F_{[ij]}=0$
in the linearized version
of eq.~(\ref{trtorsion}).  The result is
\beq
\left(D_jF^{(ij)}
-\frac{1}{2}D^i F\right)_{(1)}= \left(2\Lambda
+\frac{\tilde\alpha}{2c_5}\right) v^i
\label{v}
\eeq
One can show that
\beq
(D^jF_{(ij)})_{(1)}= \nabla^iF_{(1)(ij)}
\nonumber
\eeq
If the torsion field were absent, the left hand side of (\ref{v})
would be the linearized covariant derivative of
the  Einstein tensor and would vanish by virtue of
the contracted Bianchi identity. In the presence of torsion the Bianchi
identity is modified and is given by
\beq
D_kF_{ijlm} +
{T^n}_{kl}F_{ilmn} + {\rm cyclic}~ (klm)=0
\label{B1}
\eeq
Contracting this identity we obtain
\beq
D^{i} F_{ij} -\frac{1}{2}D_j F= {T^i}_{kj}{F^{k}}_{i} +\frac{1}{2}{T^i}_{kl}{F^{kl}}_{ji}
\label{B2}
\eeq
We note that the identitites (\ref{B1}) and  (\ref{B2}) are valid in
the full theory.
Linearizing the identity  (\ref{B2})
in the Einstein space background, we obtain
\beq
D^{i} F_{ij} -\frac{1}{2}D_j F= 2\Lambda v_j +\frac{1}{2}{T^i}_{kl}{W^{kl}}_{ji}
\label{B3}
\eeq
The comparison of (\ref{v}) and (\ref{B3}) gives
\beq
v_i= \frac{c_5}{\tilde\alpha}{T^j}_{kl}{W^{kl}}_{ij}
\nonumber
\eeq
Upon substituting for $T_{ijk}$ from (\ref{T'}) we finally obtain
\beq
v_i= \frac{4c_5}{3\tilde\alpha}W_{ijkl}t^{j[kl]}
\label{v'}
\eeq
Equation (\ref{v'}) shows that the vector field $v_i$ is not
an independently propagating field,
exactly as in the flat space. In the general Einstein space backgrounds
the field $v_i$ is determined in terms of the
tensor field $t_{ijk}$.

It is useful to note that  we can obtain a constraint on the
divergence of the field $t_{i(jk)}$ by combining (\ref{Const})
and (\ref{v'}) with the
commutators of covariant derivatives.
 We give this relation here for
later reference,
\beq 
6\nabla^i \nabla^k t_{k(ij)}=
\nabla^2v_j-\nabla_j\nabla \cdot v- \left(3\Lambda
+\frac{3\tilde\alpha}{2c_5} \right)v_j
\label{Const'}
\eeq

\subsubsection{Equation for $t_{ijk}$}

Using the symmetries of $t_{ijk}$ one can show that
\beq
[\nabla_k,\nabla_j]t^{jik}= W^{jkil}t_{lkj}
\nonumber
\eeq
This relation is instrumental in  proving that
\beq
\nabla^j\nabla^kt_{ikj}= -\frac{1}{2}\nabla^j\nabla^kt_{kji}-\frac{1}{2}W_{jkli}t^{klj}
\nonumber
\eeq
We now make use of the results of sections
\ref{sub:cons-grav}, \ref{subsub:a}, \ref{subsub:v}
and rewrite
the torsion field
equation (\ref{torsion})
as the
equation for the only
remaining component, namely, for the field $t_{ijk}$,
\beqar &&
\nabla_iF_{(1)jk}-\nabla_jF_{(1)ik}
+\frac{1}{6}(\eta_{ik}\nabla_jF_{(1)}-\eta_{jk}\nabla_iF_{(1)})
\nonumber\\ &&
-\frac{1}{3}\left(2\Lambda+
\frac{\tilde\alpha}{2c_5}\right)\{(\eta_{ik}v_j-\eta_{jk}v_i)+4t_{k[ij]}\}=0
\label{spin2}
\eeqar
Here $v_i$ should be expressed in terms of $t_{ijk}$ according
to (\ref{v'}).  It can be verified that the trace of this equation
over $j$ and $k$ as well as its divergence over $k$ are zero.  On the
other hand if we apply $\nabla^i$ to it we obtain a second order
equation
which
reads
\beqar &&(\nabla^2
-4\Lambda) F_{(1)jk}- W_{ijkl}F_{(1)}^{il}- \frac{1}{3}\left[
\nabla_j\nabla_k +\frac{1}{2}\eta_{jk}(\nabla^2 -6\Lambda)\right] F_{(1)}
\nonumber\\
&&-\frac{1}{3}\left(2\Lambda+\frac{\tilde\alpha}{2c_5}\right)\{2(\nabla_kv_j
+\nabla_jv_k) -\eta_{jk}\nabla \cdot v +6\nabla^it_{i(jk)}\}=0
\label{spin2'}
\eeqar
This equation is symmetric with respect to the interchange of
$j$ and $k$. It can be verified that it is also traceless as well as
divergence free.

\section{Symmetric spaces}
\label{sec:symmetric}

In view of complexity of eq.~(\ref{spin2}), let us specify now to
maximally symmetric backgrounds.
For these spaces,
the Weyl tensor vanishes, $W_{ijkl}=0$.
Then eq.~(\ref{v'})
implies that
\beq
v_i=0
\label{s1}
\eeq
Substitution of this in (\ref{Const}) gives
\beq
\nabla^k t_{k[mn]}=0
\nonumber
\eeq
Using (\ref{s1}) in (\ref{S}) we obtain
\beq
F_{(1)ij}= R_{(1)ij}-2\nabla^k t_{k(ij)}
\label{s3}
\eeq
The linearized Einstein equations (\ref{R}) then reduce to
\beq
c_1R_{(1)ij}= 3\tilde\alpha\nabla^kt_{k(ij)}
\label{s4}
\eeq
Let us define the symmetric tensor field $\chi_{ij}$ by
\beq
\chi_{ij}=\nabla^kt_{k(ij)}
\label{chi}
\eeq
It is immediately seen that $\chi$ is traceless and
transverse\footnote{The transversality of $\chi_{ij}$ follows from
(\ref{Const'}) when we set $v_i=0$ in that equation.}, i.e.,
\beq
\eta^{ij}\chi_{ij}=0,\quad\quad\quad \nabla^i\chi_{ij}=0
\nonumber
\eeq
Thus
$\chi_{ij}$ has only five independent components. We now write
$R_{ij}$ in (\ref{s4}) in terms of ${\chi_{ij}}$ and substitute it in
(\ref{s3}) so that $F_{(1)ij}$ is expressed in terms of
$\chi_{ij}$,
\beq
F_{(1)ij}=
\frac{3\alpha}{c_1}\chi_{ij}
\label{nov21-1}
\eeq
We insert this expression in eq.~(\ref{spin2'}) and obtain
the following Klein-Gordon type equation for the field
$\chi_{ij}$,
\beq
(\nabla^2-M_2^2)\chi_{ij}=0,
\label{s5}
\eeq
where $M_2^2$ is given by
\beq
M_2^2= 4\Lambda\left(1+\frac{c_1}{3\alpha}\right) + \frac{\tilde\alpha c_1}{3\alpha c_5}
\label{mass}
\eeq
Finally, the field $t_{k[ij]}$ is determined by eq.~(\ref{spin2}),
which by making use of (\ref{nov21-1})
is reduced to
\beq
t_{k[ij]}=
\frac{9\alpha}{4c_1(2\Lambda+\frac{\tilde\alpha}{2})}
(\nabla_i\chi_{jk}-\nabla_j\chi_{ik})
\label{t}
\eeq
As we pointed out in section \ref{sec:intro},
the tensor $t_{ijk}$ may be expressed
through its antisymmetric part  $t_{i[jk]}$, so the field $\chi_{ij}$
completely determines  $t_{ijk}$.

According to its definition (\ref{chi}), the field $\chi_{ij}$ is a 
gauge-covariant, tensor field. It follows from eqs.~(\ref{s4}),
and (\ref{t}) that both torsion field and metric
perturbations (in particular, $R_{(1) ij}$) do not vanish for this
mode. In other words, the massive tensor field results from mixing
between torsion and metric. Due to this mixing,
the massive tensor mode is sourced both by ``spin''
(source for torsion) and energy-momentum (source for metric).
This is considered in some detail in Ref.~\cite{Nikiforova:2009qr}.
Hence, the theory we discuss in this paper is a candidate for
infrared modified gravity.

Since the massive tensor mode $\chi_{ij}$ is free of pathologies
in symmetric backgrounds, its quadratic action in Miknowski
background necessarily has the Fierz--Pauli structure. The novelty
here is that the Fierz--Pauli equation is effectively deformed
into curved backgrounds, and no pathologies (say, of the Boulware--Deser
type) are introduced at least in the case of maximally symmetric
backgrounds. The explicit form of the generalized Fierz--Pauli equation
emerging in this theory is given in Ref.~\cite{Nikiforova:2009qr}
for arbitrary Einstein backgrounds.

To sum up, in the maximally symmetric background,
the only
propagating modes are  massless graviton,  massive spin zero
particle with a mass which can be read off from eq.~(\ref{aeq}) and
massive spin-2 field $\chi_{ij}$, the same as in the flat space. All
masses reduce to the  flat space values when $\Lambda=0$.

The flat space analysis indicates that for the
absence of ghosts and tachyons the parameters
should satisfy certain inequalities
which in our notations read,
\beq
c_5<0 \quad\quad\quad    c_6>0 \quad\quad\quad  \alpha<0
\quad\quad\quad  \tilde\alpha>0
\label{cond}
\eeq
Note that $c_1=M_p^2$ has to be positive.  The parameter
$c_2$ is the cosmological
term in the action and does not enter the
flat space calculations.  We see from (\ref{cc}) that
in order for $\Lambda$ to be positive $c_2$ has to
be negative.  Thus  with  positive $\Lambda$,
i.e. in  de~Sitter background,  the spin zero
field $\sigma$  in eq.~(\ref{aeq}) will be
non-tachyonic, if $4\Lambda c_5 + \tilde\alpha >0$.

It is also known that there is a unitarity bound on the
mass of the spin-2 field in de Sitter background
\cite{Higuchi:1989gz}. Comparing our notation
with that of Ref.~\cite{Higuchi:1989gz}
(especially our eq.~(\ref {s5}) with eq.~(3.10)
of Ref.~\cite{Higuchi:1989gz}) we conclude that
in our terms the untarity bound is $M_2^2>4\Lambda$.
Comparing the mass of the $\sigma$ field
with that of the spin 2 field we find  that
\beq
M_2^2= 4\Lambda + \frac{16c_6}{3\alpha c_5}M_0^2c_1
\eeq
where $M_0$ denotes the mass of the $\sigma$ field as defined by (\ref{aeq}).
Clearly
our spin-2 mass will satisfy the unitarity
bound as long as the $\sigma$ field is non-tachyonic.

\section{Conclusions}

In this paper we have examined the consistency of
the model with
a
massive spin-2 particle, whose Lagrangian
involves
quadratic terms in curvature and torsion tensors. This problem
was
analyzed a long time ago in flat space background. We have
extended this analysis to  curved background spaces. First we have
given the linearized equations for the
perturbations about arbitrary
Einstein
manifold and found that
several components of the torsion field do not propagate,
while metric perturbations correspond to massless propagating graviton.
To analyze the remaining components of the torsion field,
we then specified to maximally symmetric backgrounds.
We have
shown that, unlike in the Fierz-Pauli theory, in our model the number
and the nature of the propagating modes do not change when the
background becomes curved.
The full analysis of the propagation in the background of
an arbitrary Einstein manifold still needs to be carried out.

In maximally symmetric backgrounds, the propagating modes in our model
obey the usual Klein--Gordon type equations. Hence, there is no
superluminal propagation, again in contrast to the Fierz--Pauli theory.

There are several classes of such models which are tachyon- and ghost-free
in Minkowski background
in specific regions of their parameter space.
It will be interesting to see if there are any other
consistent subclasses among them.

\section*{Acknowledgements}
We are indebted to S. Deser for helpful correspondence.
S.R.-D. is grateful to the  organizers of the first $\Psi$G 
Workshop on the 
Consistent Modifications of Gravity 
for the invitation to present an earlier 
version of this paper  and to its participants, especially to
Diego Blas, Cedric Deffayet, Gia Dvali and 
Arkady Vainshtein for their criticism and useful comments. 
V.R. is indebted to Sergei Sibiryakov for helpful discussions.

V.P.N. has been supported in part by the National Science Foundation
grant PHY-0555620 and by a PSC-CUNY grant.
V.R.
has been supported in part by Russian Foundation for Basic
Research grant 08-02-00473.


\begin{thebibliography}{99}

\bibitem{Fierz:1939ix}
  M.~Fierz and W.~Pauli,
  Proc.\ Roy.\ Soc.\ Lond.\  A {\bf 173}, 211 (1939).

\bibitem{vanDam:1970vg}
H.~van Dam and M.~J.~G. Veltman,
\newblock Nucl. Phys. {\bf B22}, 397 (1970).

\bibitem{Zakharov:1970cc}
V.~I. Zakharov,
\newblock JETP Lett. {\bf 12}, 312 (1970).

\bibitem{Arkani-Hamed:2002sp}
N.~Arkani-Hamed, H.~Georgi and M.~D. Schwartz,
\newblock Ann. Phys. {\bf 305}, 96 (2003), [hep-th/0210184].

\bibitem{Boulware:1973my}
  D.~G.~Boulware and S.~Deser,
  Phys.\ Rev.\  D {\bf 6}, 3368 (1972).

\bibitem{Creminelli:2005qk}
P.~Creminelli, A.~Nicolis, M.~Papucci and E.~Trincherini,
\newblock JHEP {\bf 09}, 003 (2005), [hep-th/0505147].



\bibitem{Deffayet:2005ys}
C.~Deffayet and J.-W. Rombouts,
\newblock Phys. Rev. {\bf D72}, 044003 (2005), [gr-qc/0505134].


\bibitem{Osipov:2008dd}
  M.~Osipov and V.~Rubakov,
  Class.\ Quant.\ Grav.\  {\bf 25}, 235006 (2008)
  [arXiv:0805.1149 [hep-th]].

\bibitem{Deser:2001pe}
S.~Deser and A.~Waldron,
\newblock Phys. Rev. Lett. {\bf 87}, 031601 (2001), [hep-th/0102166].

\bibitem{Deser:2001wx}
S.~Deser and A.~Waldron,
\newblock Phys. Lett. {\bf B508}, 347 (2001), [hep-th/0103255].

\bibitem{Porrati:2001db}
M.~Porrati,
\newblock JHEP {\bf 04}, 058 (2002), [hep-th/0112166].

\bibitem{Rubakov:2004eb}
  V.~A.~Rubakov,
  ``Lorentz-violating graviton masses: Getting around ghosts, low strong
  coupling scale and VDVZ discontinuity,''
  arXiv:hep-th/0407104.

\bibitem{Dubovsky:2004sg}
  S.~L.~Dubovsky,
  JHEP {\bf 0410}, 076 (2004)
  [arXiv:hep-th/0409124].

\bibitem{Dubovsky:2004ud}
  S.~L.~Dubovsky, P.~G.~Tinyakov and I.~I.~Tkachev,
  Phys.\ Rev.\ Lett.\  {\bf 94}, 181102 (2005)
  [arXiv:hep-th/0411158].

\bibitem{Berezhiani:2007zf}
  Z.~Berezhiani, D.~Comelli, F.~Nesti and L.~Pilo,
  Phys.\ Rev.\ Lett.\  {\bf 99}, 131101 (2007)
  [arXiv:hep-th/0703264].


\bibitem{Rubakov:2008nh}
  V.~A.~Rubakov and P.~G.~Tinyakov,
  arXiv:0802.4379 [hep-th].

\bibitem{Hayashi:1979wj}
  K.~Hayashi and T.~Shirafuji,
  Prog.\ Theor.\ Phys.\  {\bf 64}, 866 (1980)
  [Erratum-ibid.\  {\bf 65}, 2079 (1981)].

\bibitem{Hayashi:1980ir}
  K.~Hayashi and T.~Shirafuji,
  Prog.\ Theor.\ Phys.\  {\bf 64}, 1435 (1980)
  [Erratum-ibid.\  {\bf 66}, 741 (1981)].

\bibitem{Hayashi:1980qp}
  K.~Hayashi and T.~Shirafuji,
  Prog.\ Theor.\ Phys.\  {\bf 64}, 2222 (1980).




\bibitem{Sezgin:1979zf}
  E.~Sezgin and P.~van Nieuwenhuizen,
  Phys.\ Rev.\  D {\bf 21}, 3269 (1980).

\bibitem{Hecht:1996ay}
  R.~D.~Hecht, J.~M.~Nester and V.~V.~Zhytnikov,
  Phys.\ Lett.\  A {\bf 222}, 37 (1996).

\bibitem{Yo:1999ex}
  H.~J.~Yo and J.~M.~Nester,
  Int.\ J.\ Mod.\ Phys.\  D {\bf 8}, 459 (1999)
  [arXiv:gr-qc/9902032].

\bibitem{Mao:2008gv}
  Y.~Mao,
  arXiv:0808.2063 [astro-ph].

\bibitem{Percacci:1990wy}
  R.~Percacci,
  Nucl.\ Phys.\  B {\bf 353}, 271 (1991)
  [arXiv:0712.3545 [hep-th]].

\bibitem{Nikiforova:2009qr}
  V.~Nikiforova, S.~Randjbar-Daemi and V.~Rubakov,
  arXiv:0905.3732 [hep-th].
  
\bibitem{Higuchi:1989gz}
  A.~Higuchi,
  Nucl.\ Phys.\  B {\bf 325}, 745 (1989).


\end{thebibliography}
\end{document}